\documentclass[a4paper,10pt]{article}
\usepackage[utf8x]{inputenc}
\usepackage{url,xspace,times,todonotes,color}
\usepackage{amsmath,amssymb,amsthm}
\usepackage[vlined,ruled,linesnumbered]{algorithm2e}
\usepackage{listings}

\newcommand{\ring}[1]{\mathbb{#1}}
\newcommand{\F}{\ensuremath{\ring{F}}\xspace}
\newcommand{\FZ}{\ensuremath{\ring{F}_2}\xspace}
\newcommand{\FZE}{\ensuremath{\ring{F}_{2^e}}\xspace}
\newcommand{\erange}{\ensuremath{2 \leq e \leq 10}\xspace}

\newcommand{\ord}[1]{\ensuremath{\mathcal{O}\!\left(#1\right)}}

\newcommand{\mzedt}{\texttt{mzed\_t}\xspace}
\newcommand{\mzdslicet}{\texttt{mzd\_slice\_t}\xspace}

\newcommand{\mycomputer}{2.66~Ghz Intel i7\xspace}

\newtheorem{definition}{Definition}
\newtheorem{lemma}{Lemma}

\title{The M4RIE library for dense linear algebra over small fields with even characteristic}
\author{Martin R.\ Albrecht\\\url{malb@lip6.fr}}

\renewcommand{\todo}[1]{}

\begin{document}

\maketitle

\section{Introduction}
\label{sec:introduction}

Linear algebra over small finite fields has many direct applications, such as cryptography and coding theory. Other applications include efficient linear algebra over the rationals, e.g., by reducing such computations to a series of computations modulo small primes, and solving non-linear systems of equations using Gröbner bases \cite{lazard:eurocal83}. For the latter recent work has emphasised that the right choice of linear algebra algorithms and implementations can make a significant impact on the performance of Gröbner basis algorithms \cite{faugere-lachatre:pasco2010}. 

Compared to other small finite fields, those with even characteristic have some special properties which make them a prominent choice for designing cryptographic and coding systems -- cf., the AES \cite{Daemen2002} as a prime example. For instance, addition is simply XOR and is hence natively available on modern CPUs, the same cannot be said for other small finite fields. Yet, this family of finite fields has not received much attention in the literature on linear algebra. However, that the current state of the art in the literature leaves something to be desired can be observed from the following simple benchmark: Computing the (reduced) row echelon form of a random $4,000 \times 4,000$ matrix over $\F_{4}$ on a \mycomputer CPU takes 5s using GAP 4.4.12 \cite{GAP4} (not reduced) or even 876.6s and 2805.8s  using NTL 5.4.2 \cite{NTL}  (not reduced) and Sage \cite{sage} (reduced) respectively. Furthermore, LinBox's FFPACK \cite{FFPACK} performs \todo{get timings over \FZE} the same task over the prime field $\F_{3}$ (reduced) in 4.55s. For comparision, the closed source system Magma \cite{magma} can compute the reduced row echelon form of a dense $4,000 \times 4,000$ matrix over $\F_{4}$ in 0.64s (reduced) and the same operation over $\F_2$ takes 0.054s using the M4RI library \cite{m4ri-ple}.

In this work, we present the M4RIE library which implements efficient algorithms for linear algebra with dense matrices over \FZE for \erange.\footnote{Future versions will support $e \leq 16$.} As the name of the library indicates, it makes heavy use of the M4RI library \cite{m4ri} both directly (i.e., by calling it) and indirectly (i.e., by using its concepts). The contributions of this work are as follows. 
We provide an open-source GPLv2+ C library for efficient linear algebra over \FZE with \erange. In this library we implemented an idea due to Bradshaw and Boothby \cite{BB09} which reduces matrix multiplication over $\F_{p^n}$ to a series of matrix multiplications over $\F_p$. Furthermore, we propose a caching technique -- \emph{Newton-John} tables -- to avoid finite field multiplications which is inspired by Kronrod's method (``M4RM'') \cite{ADKF70,matmulgf2} for matrix multiplication over $\F_2$. Using these two techniques we provide asymptotically fast triangular solving with matrices (TRSM) and PLE-based \cite{jeannerod-pernet-storjohann:ple2010} Gaussian elimination. As a result, we are able to significantly improve upon the state of the art in dense linear algebra over \FZE with \erange. For example, the above benchmark is completed in 0.4s by our library.

\section{Notation} \label{sec:notation}
We represent elements in $\FZE \cong \FZ[x]/\langle f \rangle$, with $f \in \FZ[x]$, $\deg(f) = e$ and $f$ irreducible, as polynomials $\sum_{i=0}^{e-1} a_i x^i$ or as coefficient vectors $(a_{e-1},\dots,a_0)$ where $a_i \in \FZ$. We sometimes identify the coefficient vector $(a_{e-1},\dots,a_0)$ with the integer $\sum_{i=0}^{e-1} a_i2^i$, e.g., when indexing tables by finite field elements. By $\alpha$ we denote some root of the primitive polynomial $f$ of \FZE. By $A_i$ we denote the $i$-th row of the matrix $A$ and by $A_{i,j}$ the entry in row $i$ and column $j$ of $A$. We start counting at zero. We represent permutation matrices as LAPACK-style permutation vectors. That is to say, that for example the permutation matrix 
$$\left[ \begin{array}{ccc}
1 & 0 & 0 \\
0 & 0 & 1 \\
0 & 1 & 0 \\
\end{array} \right]$$
is stored as $P=[0, 2, 2]$, where for each index $i$ the entry $P_i$ encodes which (row or column) swap should to be performed on the input matrix. This allows to apply permutations in-place.

\section{Matrix representation} \label{sec:representation}

The M4RIE library features two matrix types, each of which is optimised for certain operations. Both representations use one or more M4RI matrices as data storage and hence re-use M4RI's matrix window concept \cite{matmulgf2}, allocation routines and data structures.

\subsection{packed: \mzedt}  \label{sec:bitpacked}

Considering the polynomial representation of elements in \FZE, we may bit-pack several such elements in one machine word. Since we are using the M4RI library as actual data storage, this means that words hold 64 bits \cite{matmulgf2}. Hence, the \mzedt data type packs elements of \FZE in 64-bit words. However, instead of packing as many elements as possible into one word, every element is padded to the next length dividing 64. Thus, for example, elements in $\F_{32}$ are represented as polynomials of degree $8$ where the top three coefficients are always zero. While this wastes some storage space and CPU time, it allows for more compact code by reducing what cases have to be considered. The second row of Figure~\ref{fig:example} gives an example.

In this representation additions are very cheap since we can disregard any element boundaries and simply call M4RI's addition routines. Scalar multiplication, on the other hand, is much more expensive. Either we perform a table look-up for each \emph{element} or we perform bit operations on words which perform multiplication and modular reduction in parallel on all elements of a word. In either case, multiplication is considerably more expensive than addition.

\subsection{sliced: \mzdslicet} \label{sec:bitsliced}

Instead of representing matrices over \FZE as matrices over polynomials we may represent them as polynomials with matrix coefficients. That is, for each degree we store matrices over \FZ which hold the coefficients for this degree. Hence,  the data type \mzdslicet for matrices over $\FZE$ internally stores $e$-tuples of M4RI matrices, i.e., matrices over \FZ. We call each M4RI matrix for some degree $i$ a \emph{slice} and refer to the operation converting from \mzedt to \mzdslicet as \emph{slicing}. The inverse operation is called \emph{clinging}. The third row of Figure~\ref{fig:example} gives an example of the \mzdslicet representation.

Addition is performed by adding each slice independently and hence is quite efficient. Scalar multiplication, on the other hand, has to rely on similar techniques as in \mzedt. However, in addition, data locality in \mzdslicet is worse than in \mzedt. Thus, here too, scalar multiplication is much more expensive than addition.

Thus, in this work, we present algorithms for matrix multiplication and elimination where we avoid many scalar multiplications.

\begin{figure}[ht]
\begin{eqnarray*}
A =& \left(\begin{array}{cc}
		  \alpha^2 + 1 & \alpha \\
		  \alpha + \alpha + 1 & 1 \\
       \end{array}\right)\\
 = & \left[\begin{array}{cc}\Box101&\Box010\\
                                  \Box011&\Box001\\
     \end{array}\right]\\
 = & \left(\left[\begin{array}{cccc}
1&0\\
1&1\\
\end{array}\right], \left[\begin{array}{cc}
0&1\\
1&0\\
\end{array}\right], \left[\begin{array}{cc}
1&0\\
0&0\\
\end{array}\right]\right)
\end{eqnarray*}
\caption{$2 \times 2$ matrix over $\F_{8}$}
\label{fig:example}
\end{figure}

\section{Newton-John tables} \label{sec:newton-john}

To explain the main idea behind Newton-John tables, consider matrix multiplication as given in Algorithm~\ref{alg:cubic_mat_mul}.

\begin{algorithm}[H]
\KwIn{$A$ -- $m \times \ell$ matrix}
\KwIn{$B$ -- $\ell \times n$ matrix}
\KwOut{$C = A \cdot B$}
\Begin{
\For{$0 \leq i < m$}{
  \For{$0 \leq j < \ell$}{
 $C_j \gets C_j + A_{j,i} \times B_i$\nllabel{alg:cubic_mat_mul:work}\;
}
}
\Return{$C$}\;
}
\caption{Cubic matrix multiplication}
\label{alg:cubic_mat_mul}
\end{algorithm}

This algorithm uses $m\cdot \ell \cdot n$ finite field multiplications and the same number of additions. That is, in line~\ref{alg:cubic_mat_mul:work} the row $B_j$ is scaled by $A_{j,i}$ and then added to the row $C_j$. Observe that $B_j$ is rescaled $\ell$-times, while there are $2^e$ different values for $A_{j,i}$ and hence multiples of $B_j$. Indeed, if $2^e < \ell$ it is advantageous to precompute all possible $2^e$ multiples of $B_j$ and  to store these multiples in a table indexed by finite field elements. These precompuation tables are quite similar to Kronrod's method for matrix multiplication, also sometimes referred to as ``greasing''. Hence, we call these tables \emph{Newton-John tables} to honour Olivia Newton-John's work in \cite{grease}. 

We also note that we create these tables in less than $2^e$ multiplications. That is, we first compute $\alpha^i \cdot B_j$ for all $0 \leq i < e$. Then, we compute each multiple of $B_j$ as a linear combination of $(\alpha^0\cdot B_j,\dots \alpha^{e-1}\cdot B_j)$ which we just computed. Using Gray codes for the addition step we can thus construct all $2^e$ multiples using $2^e$ additions \cite{graycode}. The subroutine creating these tables is given in Algorithm~\ref{alg:maketable}. 

The complete algorithm is given in Algorithm~\ref{alg:travolta_mat_mul} which costs $m \cdot (2^e + \ell) \cdot n$ additions and $m \cdot e \cdot n$ multiplications. Note that $e$ is a constant here and asymptotically we thus achieve $\ord{n^3}$ additions and $\ord{n^2}$ multiplications.

\begin{algorithm}[H]
\KwIn{$B$ -- an $1 \times n$ matrix}
\KwOut{$T$ -- a $2^e \times n$ matrix with each row a multiple of $B$}
\Begin{
$M \gets e \times n$ matrix\;
$T \gets 2^e \times n$ matrix\;
\For{$0 \leq k < e$}{
  $M_k \gets \alpha^k \cdot B$\;
}
$T \gets$ all linear combinations of rows of $M$\; 
}
\Return{$T$};
\caption{\textsc{MakeTable}}
\label{alg:maketable}
\end{algorithm}

\begin{algorithm}[H]
\KwIn{$A$ -- $m \times \ell$ matrix}
\KwIn{$B$ -- $\ell \times n$ matrix}
\KwOut{$C = A \cdot B$}
\Begin{
\For{$0 \leq i < m$}{
  $T \gets$ \textsc{MakeTable}($B_i$)\; 
  \For{$0 \leq j < \ell$}{
  $x \gets A_{j,i}$ as an integer\;
  $C_j \gets C_j + T_x$\nllabel{alg:travolta_mat_mul:inner}\;
}
}
\Return{$C$}\;
}
\caption{Newton-John multiplication}
\label{alg:travolta_mat_mul}
\end{algorithm}

Many variants of this basic algorithm are possible. For instance, we may use more than one Newton-John table or process the data in blocks for better cache friendliness (cf., \cite{matmulgf2} for both techniques). Furthermore, if $2^e$ is too big to precompute $T$ we may precompute only $M$ (cf., Algorithm~\ref{alg:maketable}) and perform $e$ additions in line~\ref{alg:travolta_mat_mul:inner}.
Our library uses eight Newton-John tables and processes matrices in blocks that fit into L2 cache. Since $e$ is small we always compute the full table $T$.

Of course, this algorithm is not asymptotically fast. Hence, we only use it as a base case for the Strassen-Winograd algorithm \cite{Strassen} for matrix multiplication which has complexity $\ord{n^{\log_2 7}}$. In our implementation we cross over to the base case roughly when the submatrices fit into L2 cache; however, the exact value depends on the size of the finite field. Table~\ref{tab:travolta_mat_mul_times} lists CPU times multiplying two $1,000 \times 1,000$ matrices in our implementation of Strassen-Winograd on top of Newton-John multiplication (abbreviated S-W/N-J in the following), in Magma and GAP \todo{LinBox?} (cf., Appendix~\ref{app:others} for a brief discussion of Magma's and GAP's implementations). Note that the hex string in the header of the last column indicates which revision of the public source code repository\footnote{cf., \url{https://bitbucket.org/m4rie}.} was used to produce these times.

\begin{table}[ht]
\begin{small}
\begin{center}
\begin{tabular}{|r|r|r|r|}
\hline
 $e$ & Magma & GAP & S-W/N-J \\
     & {\footnotesize 2.15-10} & {\footnotesize 4.4.12} & {\footnotesize 6b24b839a46f}\\
\hline
 2 & 0.013s &  0.216s & 0.012s\\
 3 & 0.036s &  0.592s & 0.020s\\
 4 & 0.074s &  0.588s & 0.022s\\
 5 & 1.276s &  1.568s & 0.048s\\
 6 & 1.286s &  1.356s & 0.059s\\
 7 & 1.316s &  1.276s & 0.082s\\
 8 & 1.842s &  1.328s & 0.160s\\
 9 & 3.985s & 64.700s & 0.626s\\
10 & 4.160s & 59.131s & 1.080s\\
\hline
\end{tabular} 
\end{center}
\end{small}
\caption{Multiplication of $1,000 \times 1,000$ matrices on \mycomputer}
\label{tab:travolta_mat_mul_times}
\end{table}

\subsection{Gaussian elimination} \label{sec:gauss}

Newton-John tables can also be used in Gaussian elimination, as shown in Algorithm~\ref{alg:travolta_gauss}. This algorithm uses $r \cdot (n + 2^e) \cdot n$ additions and $r \cdot (e+1) \cdot n$ multiplications, which gives an asymptotic complexity of $\ord{n^3}$ additions and $\ord{n^2}$ multiplications. Again, a variety of variants are possible such as multiple Newton-John tables (similar to \cite{matmulgf2}). Our implementation uses six Newton-John tables.

\begin{algorithm}[H]
\KwIn{$A$ -- $m \times n$ matrix}
\KwOut{$r$ -- the rank of $A$}
\KwResult{$A$ is in reduced row echelon form}
\SetKw{KwContinue}{continue}
\SetKw{KwBreak}{break}
\Begin{
$r \gets 0$\;
\For{$0 \leq j < n$} {
  \For{$s \leq i < m$} {  
      \If{$A_{i,j} \neq 0$}{
        $A_{i} \gets A_{i,j}^{-1} \cdot A_{i}$\nllabel{alg:travolta_gauss:pivot}\;
        swap the rows $i$ and $r$ in $A$\;
		$T \gets$ \textsc{MakeTable}($A_r$)\; 
		\For{$0 \leq k < m$} {
		  \lIf{$k = s$}{\KwContinue\;}
		  $x \gets A_{k,j}$ as an integer\;
		  $A_k \gets A_k + T_x$\nllabel{alg:travolta_gauss:add}\;
		}
        $r \gets r + 1$\;
        \KwBreak;
	}
  }
}
\Return $r$\nllabel{alg:travolta_gauss:compress}\;
}
\caption{Newton-John Gauss elimination}
\label{alg:travolta_gauss}
\end{algorithm}

In Table~\ref{tab:travolta_elim_times} we give CPU times for computing the reduced row echelon form of random $1,000 \times 1,000$ matrices over \FZE in Magma, GAP \todo{LinBox} and our implementation. Note that GAP's \texttt{SemiEchelonMat} command does not compute the \emph{reduced} row echelon form. Hence, to normalise the data we multiplied all GAP timings in Table~\ref{tab:travolta_elim_times} by two.

\begin{table}[ht]
\begin{small}
\begin{center}
\begin{tabular}{|r|r|r|r|}
\hline
 $e$ & Magma & GAP & Newton-John \\
     & {\footnotesize 2.15-10} & {\footnotesize 4.4.12} & {\footnotesize 6b24b839a46f}\\
\hline
 2 & 0.028s &  0.184s & 0.012s\\
 3 & 0.045s &  0.496s & 0.019s\\
 4 & 0.054s &  0.560s & 0.022s\\
 5 & 0.690s &  1.224s & 0.042s\\
 6 & 0.670s &  1.168s & 0.048s\\
 7 & 0.700s &  1.104s & 0.060s\\
 8 & 0.866s &  1.136s & 0.081s\\
 9 & 1.523s & 35.634s & 0.427s\\
10 & 1.540s & 36.154s & 0.831s\\
\hline
\end{tabular} 
\end{center}
\end{small}
\caption{Elimination of $1,000 \times 1,000$ matrices on \mycomputer}
\label{tab:travolta_elim_times}
\end{table}

\subsection{PLE decomposition} \label{sec:ple}

Algorithm~\ref{alg:travolta_gauss} can be modified to compute the PLE decomposition instead of the row echelon form. Since this definition is lesser well-known we reproduce it below. For a more detailed treatment of PLE decomposition see \cite{jeannerod-pernet-storjohann:ple2010}.

\begin{definition}[PLE]
Let $A$ be a $m\times n$ matrix over a field $K$. A PLE decomposition of $A$ is a triple of matrices $P,L$ and $E$ such that $P$ is a $m\times m$ permutation matrix, $L$ is a unit lower triangular matrix, and $E$ is a $m\times n$ matrix in row-echelon form, and $A=PLE$.
\end{definition}

\begin{lemma}[\cite{jeannerod-pernet-storjohann:ple2010}]
A PLE decomposition of any $m \times n$ matrix $A$ can be stored in-place in $A$.
\end{lemma}

For the sake of simplicity, we compute a minor variant of PLE in our library. That is, $L$ is not necessarily unit lower triangular, i.e., we do rescale the pivot row get leading entry $1$. Then, the changes necessary for Algorithm~\ref{alg:travolta_gauss} to compute this variant of PLE decomposition (in-place) are:
\begin{itemize}
 \item Store $i$ and $j$ in two vectors $P$ and $Q$ in line~\ref{alg:travolta_gauss:pivot};
 \item only start addition in column $c+1$ in line~\ref{alg:travolta_gauss:add} in order to keep $L$ in place;
 \item perform column swaps below and on the main diagonal right before line~\ref{alg:travolta_gauss:compress} to compress $L$.
\end{itemize}

An alternative perspective on  Newton-John-table based PLE decomposition is to consider it as a block iterative PLE decomposition (cf., \cite{m4ri-ple}) with Newton-John table based multiplication updates to the right hand side.  

Note, this algorithm is not asymptotically fast, hence its main application is as a base case for asymptotically fast PLE decomposition \cite{jeannerod-pernet-storjohann:ple2010} which reduces PLE decomposition to matrix multiplication. For this, one last building block is needed:

\subsection{TRiangular Solving with Matrices} \label{sec:trsm}

Triangular system solving with matrices can also be achieved using Newton-John tables. As a example, we given an algorithm for solving $X = U^{-1} \cdot B$ with $U$ upper triangular in Algorithm~\ref{alg:travolta_trsm_upper_left}. We note that Algorithm~\ref{alg:travolta_trsm_upper_left} is essentially block iterative TRSM with Newton-John table based multiplication. Yet, we present it here for completeness.

\begin{algorithm}[H]
\KwIn{$U$ -- $m \times m$ upper triangular matrix}
\KwIn{$B$ -- $m \times n$ matrix}
\KwResult{$X = U^{-1} \cdot B$ is stored in $B$}
\SetKw{KwContinue}{continue}
\SetKw{KwBreak}{break}
\Begin{
\For{$m > i \geq 0$}{
    $B_i \gets U_{i,i}^{-1} \cdot B_i$\;
	$T \gets$ \textsc{MakeTable}($B_i$)\; 
	\For{$0 \leq j < i$} {
	  $x \gets U_{j,i}$\;
	  $B_j \gets B_j + T_x$\;
	}
}
}
\caption{Newton-John TRSM upper left}
\label{alg:travolta_trsm_upper_left}
\end{algorithm}

\section{Karatsuba multiplication} \label{sec:karatsuba}

Recall that \mzdslicet represents matrices over \FZE as polynomials with matrices over \FZ as coefficients. Using this representation, matrix multiplication then can be accomplished by performing polynomial multiplication and subsequent modular reduction. For example, assume we want to compute $C=A\cdot B$ where $A$ and $B$ are over $\F_{2^2}$. We rewrite $A$ as $ A_1x + A_0$ and $B$ as $ B_1x + B_0$, the product is then ${\tilde C} = A_1B_1x^2 + (A_1B_0 + A_0B_1)x + A_0B_0$ which reduces to $C = (A_1B_1 + A_1B_0 + A_0B_1)x + A_0B_0 + A_1B_1$ modulo the primitive polynomial $f = x^2 + x + 1$ of $\F_{2^2}$. Hence, matrix multiplication over \FZE can be reduced to matrix multiplication and addition over \FZ. Using naive polynomial arithmetic we get that matrix multiplication over \FZE costs $e^2$ matrix multiplications over \FZ. However, using Karatsuba polynomial multiplication we can reduce this to $e^{\log_2 3} \approx e^{1.584}$. To get back to the above example, we can rewrite it as $C = ((A_1 + A_0)(B_1 + B_0) + A_0B_0)x + A_0B_0 + A_1B_1$ and hence multiplication costs $3$ instead of $4$ multiplications over \FZ. This was first explicitly proposed for matrices over $\F_{p^n}$ by Bradshaw and Boothby in \cite{BB09}. However, this technique has been used for linear algebra over $\F_{2^k}$ with $2 \leq k \leq 4$ in Magma for some time \cite{steel:gf2k}.

Concrete costs are given in Table~\ref{tab:karatsuba_mat_mul_times} where the first column lists the CPU time for multiplying two $4,000 \times 4,000$ matrices using Strassen-Winograd on top of Newton-John multiplication, the column ``$M$'' indicates how many $4,000 \times 4,000$ matrix multiplications over \FZ can be achieved in the same time using the M4RI library, the column ``naive'' lists how many multiplications would be needed by naive polynomial multiplication, the column ``\cite{M05}'' lists the best known complexity for Karatsuba-like formulas, the last column shows the number of multiplications which our Karatsuba-like implementation actually achieves. The absolute time of our bitsliced implementation is given in the column ``Bitslice''. Table~\ref{tab:karatsuba_mat_mul_times} also compares our implementation with the previous two best implementations GAP and Magma \todo{LinBox?}.

However, we note that Karatsuba based multiplication needs more memory than Strassen on top of Newton-John multiplication. Our implementation uses three temporary matrices over \FZ. We finish this section by pointing out in principle more efficient polynomial multiplication algorithms for $\FZ[x]$ can be applied (cf., \cite{Brent:2008:FMG:1789715.1789728}). However, due to the small degrees considered in this work it does not seem advantageous.

\begin{table}[ht]
\begin{small}
\begin{center}
\begin{tabular}{|r||r|r||r|r|r|r|r||r|}
\hline
 $e$ & Magma & GAP & S-W/N-J & Bitslice	& $M$ &	naive	 & \cite{M05} & Bitslice/\\
     & {\footnotesize 2.15-10} & {\footnotesize 4.4.12} & & & & &  & M4RI\\
\hline
 2 &   1.220s & 12.501s &  0.630s & 0.224s &   8.8 &  4 &  3 &  3.1\\ 
 3 &   2.020s & 35.986s &  1.480s & 0.448s &  20.8 &  9 &  6 &  6.3\\ 
 4 &   5.630s & 39.330s &  1.644s & 0.693s &  23.1 & 16 &  9 &  9.7\\ 
 5 &  94.740s & 86.517s &  3.766s & 1.005s &  53.0 & 25 & 13 & 14.2\\ 
 6 &  89.800s & 85.525s &  4.339s & 1.336s &  61.1 & 36 & 17 & 18.8\\ 
 7 &  82.770s & 83.597s &  6.627s & 1.639s &  93.3 & 49 & 22 & 23.1\\ 
 8 & 104.680s & 83.802s & 10.170s & 2.140s & 143.2 & 64 & 27 & 30.1\\ 
\hline
\end{tabular}
\caption{Multiplication of $4,000 \times 4,000$ matrices over $\F_{2^e}$ on \mycomputer.}
\label{tab:karatsuba_mat_mul_times}
\end{center}
\end{small}
\end{table}

\section{Echelon Forms} \label{sec:echelon}

Putting these building blocks together
\begin{itemize}
 \item[(1)] Karatsuba multiplication, 
 \item[(2)] Newton-John-based PLE decomposition,
 \item[(3)] asymp\-totic\-ally-fast PLE decomposition,
 \item[(4)] Newton-John-based Triangular Solving with Matrices (TRSM) and 
 \item[(5)] asymptotically-fast TRSM,
\end{itemize}
we can construct asymptotically fast Gaussian elimination, e.g., computation of (reduced) row echelon forms (cf., \cite{jeannerod-pernet-storjohann:ple2010}). Our implementation uses \mzdslicet as representation for large matrices and switches over to \mzedt when the submatrix currently considered fits into L2 cache. Table~\ref{tab:echelonform} lists CPU times for computing the (reduced) row echelon form using Magma (reduced), GAP (not reduced) \todo{LinBox?} and our implementation (reduced). Note that our implementation as of now only implements asymptotically fast PLE decomposition up to $e=8$, for $e\in\{9,10\}$ Newton-John table based Gaussian elimination is used.

\begin{table}[ht]
\begin{small}
\begin{center}
\begin{tabular}{|r|r|r|r|}
\hline
 $e$ & Magma & GAP & M4RIE \\
     & {\footnotesize 2.15-10} & {\footnotesize 4.4.12} & {\footnotesize 6b24b839a46f}\\
\hline
  2 &    6.040s & 162.658s &   3.310s\\
  3 &   14.470s & 442.522s &   5.332s\\
  4 &   60.370s & 502.672s &   6.330s\\
  5 &  659.030s &      N/A &  10.511s\\
  6 &  685.460s &      N/A &  13.078s\\
  7 &  671.880s &      N/A &  17.285s\\
  8 &  840.220s &      N/A &  20.247s\\
\hline
  9 & 1630.380s &      N/A & 260.774s\\
 10 & 1631.350s &      N/A & 291.298s\\
\hline
\end{tabular} 
\end{center}
\end{small}
\caption{Elimination of $10,000 \times 10,000$ matrices on \mycomputer}
\label{tab:echelonform}
\end{table}

\bibliographystyle{plain}
\bibliography{../literature}

\appendix
\section{Other implementations} \label{app:others}

\begin{description}
 \item[GAP] \cite{GAP4} packs finite field elements of size $2 < s \leq 2^8$ into words using 8 bits per entry. Arithmetic is implemented using table look ups.  Multiplication is performed using cubic matrix multiplication. Row echelon forms are computed using cubic Gaussian elimination.
 \item[LinBox/FFPACK] \cite{FFPACK} uses floating point numbers to represent finite field elements. For extension fields, elements are represented as ``sparse'' integers, such that there are enough zeroes between two coefficients to avoid the carry travelling too far. However, this feature is not readily exposed to the end-user and requires some tweaking to work. FFPACK implements Strassen-Winograd multiplication and asymptotically fast PLUQ decomposition for Gaussian elimination.
 \item[Magma] \cite{magma} implements asymptotically fast matrix multiplication and reduces Gaussian elimination to LQUP decomposition. For $\F_{2^k}$ with $2 \leq k \leq 4$ a bitsliced representation similar to our \mzdslicet is used in combination with Karatsuba-like formulas for polynomial multiplication. For $5 \leq k \leq 20$ elements in $\F_{2^k}$ are represented using Zech logarithms. For larger $k$ a packed polynomial representation is used similar to our \mzedt \cite{steel:gf2k}.
\end{description}

\end{document}